\documentclass[a4paper,10pt]{revtex4}
\usepackage{mathrsfs}
\usepackage{graphicx}
\usepackage{latexsym}
\usepackage{amsmath}
\usepackage{amssymb}
\usepackage{textcomp}
\usepackage{amsbsy}
\usepackage{graphics}
\usepackage{epstopdf}
\usepackage{color}

\allowdisplaybreaks[4]

\begin{document}

\tolerance=5000

\title{Extended matter bounce scenario in ghost free $f(R,\mathcal{G})$ gravity compatible with GW170817}

\author{E.~Elizalde,$^{1}$\,\thanks{elizalde@ieec.uab.es}
S.~D.~Odintsov,$^{2,3,4}$\,\thanks{odintsov@ieec.uab.es}
V.~K.~Oikonomou,$^{5,6,7}$\,\thanks{v.k.oikonomou1979@gmail.com}
Tanmoy~Paul$^{8,9}$\,\thanks{pul.tnmy9@gmail.com}} \affiliation{ $^{1)}$ Institut de Ci\`encies de l'Espai (ICE-CSIC/IEEC),\\
Campus UAB, c. Can Magrans s/n, 08193, Barcelona, Spain. \\
$^{2)}$ ICREA, Passeig Luis Companys, 23, 08010 Barcelona, Spain\\
$^{3)}$ Institute of Space Sciences (IEEC-CSIC) C. Can Magrans
s/n,
08193 Barcelona, Spain\\
$^{4)}$ Institute of Physics, Kazan Federal University, 420008 Kazan, Russia\\
$^{5)}$ Department of Physics, Aristotle University of
Thessaloniki, Thessaloniki 54124,
Greece\\
$^{6)}$ International Laboratory for Theoretical Cosmology, Tomsk
State University of Control Systems
and Radioelectronics (TUSUR), 634050 Tomsk, Russia\\
$^{7)}$ Tomsk State Pedagogical University, 634061 Tomsk, Russia\\
$^{8)}$ Department of Physics, Chandernagore College, Hooghly - 712 136.\\
$^{(9)}$ Department of Theoretical Physics,\\
Indian Association for the Cultivation of Science,\\
2A $\&$ 2B Raja S.C. Mullick Road,\\
Kolkata - 700 032, India }

\tolerance=5000

\begin{abstract}

In the context of a ghost free $f(R,\mathcal{G})$ model, an
extended matter bounce scenario is studied where the form of the
scale factor is given by $a(t) = (a_0t^2 + 1)^n$. The ghost free
character of the model is ensured by the presence of a Lagrange
multiplier, as developed in \cite{Nojiri:2018ouv}. The conditions
under which, in this model, the speed of gravitational waves
becomes equal to the speed of light (equal to one, in natural
units), thus becoming compatible with the striking event GW170817,
is investigated. It is shown that  this happens for a class of
Gauss-Bonnet (GB) coupling functions ($h(t)$) which satisfies a
constraint equation of the form $\ddot{h} = \dot{h}H$, with $H$
the Hubble parameter. This constraint is then imposed on the ghost
free $f(R,\mathcal{G})$ gravity theory to be consistent with the
GW170817 event, subsequently, the corresponding non-singular
bouncing cosmology with the aforementioned scale factor is
extensively studied. The forms of the coupling function and
Lagrange multiplier in the ``low curvature limit'' of the theory
are reconstructed, yielding a viable approximation for $n < 1/2$.
Correspondingly, by solving the cosmological perturbation
equation, the main observable quantities, namely the spectral
index, tensor to scalar ratio, and the running index are
determined and confronted with the latest Planck 2018 data.
Consistency with the data is proven for those parametric regimes
that which correspond to $n < 1/2$. This makes the low curvature
approximation a viable one for calculating the scalar and tensor
power spectra.
\end{abstract}


\maketitle
\section{Introduction}

A major issue in theoretical cosmology is to ascertain whether the
Universe was created from an initial singular or if, on the
contrary, the Universe initiated its expansion from a
non-singular, bounce-like stage. In other words, if the Universe
evolved according to the standard cosmology evolution, for which
the Universe emerges from an initial spacelike singularity, or if,
in accordance with the bouncing theory, the Universe started its
expansion with a non-zero value of the scalae factor. The
inflationary picture
\cite{guth,Linde:2005ht,Langlois:2004de,Riotto:2002yw,nb1,barrow1,barrow2,Banerjee:2017lxi,Chakraborty:2018scm,
Elizalde:2018now,Elizalde:2018rmz,Nojiri:2019dwl} has earned a lot
of support, since it is able to resolve the horizon and flatness
problems, and also generate an almost scale invariant power
spectrum which is perfectly consistent with the observational
data. However even the most recent observations could not yet
undoubtedly confirm that inflation indeed took place, a direct
proof of this fact (as would be, e.g., the direct or indirect
detection of associated primordial gravitational waves) is still
missing.

One of the alternative theories different from inflation is the
bouncing scenario
\cite{Brandenberger:2012zb,Brandenberger:2016vhg,Battefeld:2014uga,Novello:2008ra,Cai:2014bea,deHaro:2015wda,Nojiri:2019lqw,Lehners:2011kr,Lehners:2008vx,
Cheung:2016wik,Cai:2016hea,Cattoen:2005dx,Li:2014era,Brizuela:2009nk,Cai:2013kja,Quintin:2014oea,Cai:2013vm,Poplawski:2011jz,
Koehn:2015vvy,Odintsov:2015zza,Nojiri:2016ygo,Oikonomou:2015qha,Odintsov:2015ynk,Koehn:2013upa,Battarra:2014kga,Martin:2001ue,Khoury:2001wf,
Buchbinder:2007ad,Brown:2004cs,Hackworth:2004xb,Nojiri:2006ww,Johnson:2011aa,Peter:2002cn,Gasperini:2003pb,Creminelli:2004jg,Lehners:2015mra,
Mielczarek:2010ga,Lehners:2013cka,Cai:2014xxa,Cai:2007qw,Cai:2010zma,Avelino:2012ue,Barrow:2004ad,Haro:2015zda,Elizalde:2014uba,Das:2017jrl}.
This theory is also  able to generate a nearly scale invariant
power spectrum, thus becoming compatible with the observational
data available. Moreover, a definite advantage of the bouncing
scenario is that it avoids any spacetime singularity. In fact, the
Big-Bang singularity could just be a manifestation of the
shortcomings of classical gravity, which is certainly unable to
describe the physical evolution at such small scales. Quite
possibly, a yet-to-be-built quantum theory of gravity might be
able to resolve the Big Bang singularity, as turned out to be
actually the case in classical electrodynamics, with the Coulomb
potential singularities at the origin of the potential, which are
resolved in the context of quantum electrodynamics. However, in
the absence of a fully accepted quantum gravity theory, the
bouncing scenario reveals as a most promising theory  to deal with
this issue.

Among the various bouncing models proposed so far, the matter
bounce scenario (MBS)
\cite{deHaro:2015wda,Cai:2008qw,Finelli:2001sr,Quintin:2014oea,Cai:2011ci,
Haro:2015zta,Cai:2011zx,Cai:2013kja,
Haro:2014wha,Brandenberger:2009yt,deHaro:2014kxa,Odintsov:2014gea,
Qiu:2010ch,Oikonomou:2014jua,Bamba:2012ka,deHaro:2012xj,
WilsonEwing:2012pu} has attracted a lot of attention, since it
produces a nearly scale invariant power spectrum. In the matter
bounce theory, the Universe evolved from an epoch at large
negative time in the contracting era where the primordial
spacetime perturbations are generated deeply inside the Hubble
radius, and is thus able to solve the horizon problem. After
bouncing, the Universe starts to expand with an evolution that is
symmetric to the contraction phase. Here, it may be mentioned that
another popular bounce scenario is the Big-Bounce one, which is
also able to give a nearly scale invariant power spectrum. However
in this last scenario, the perturbations are produced near the
bouncing point, unlike what happens in the matter bounce model,
where the perturbations originate, as mentioned earlier, deeply in
the contracting phase. This is the scenario we will consider in
the present paper, in particular the extended matter bounce
scenario about which we will discuss later.

Despite the mentioned successes, the MBS still faces some
problems, firstly, in an exact MBS characterized by a single
scalar field, the power spectrum turns out to be exactly scale
invariant (i.e the spectral index of the curvature perturbation is
exactly equal to one), what does not match the observational
constraints. Such inconsistency was also confirmed in
\cite{Odintsov:2014gea} from a slightly different viewpoint,
namely from an $F(R)$ gravity theory. It turns out that $F(R)$
models can be equivalently mapped to scalar-tensor ones via
conformal transformation of the metric
\cite{Das:2017htt,Elizalde:2018rmz,Elizalde:2018now} and, thus,
the inconsistencies of the spectral index in the two different
models are well justified. Secondly, according to the Planck 2018
data, the running of the spectral index is constrained to be
$-0.0085 \pm 0.0073$. However, for the MBS in the case of a single
scalar field model, the running of the index becomes zero and
hence it is not compatible with observations. A zero running index
is a consequence of the first problem. If the spectral index
becomes exactly scale invariant, then the variation of the
spectral index with respect to the mode momentum, i.e the running
index, will obviously vanish. Thirdly, in the simplest MBS model,
the amplitude of scalar fluctuations is found to be comparable to
that of tensor perturbations, which in turn makes the value of the
tensor-to-scalar ratio to be of order one, again in conflict with
the Planck constraints. Lastly, the null energy condition in most
of the bouncing models is violated when the bouncing is realized.

In order to solve all these issues, theories of modified gravity
\cite{Nojiri:2010wj,Nojiri:2017ncd} have been added to the
picture, with success in many of the cases. For instance, in the
so-called quasi-matter bounce scenario (which improves the exact
matter bounce one), according to which the scale factor of the
Universe evolves as $t^{3(1+w)}$ (with $w \neq 0$), deeply into
the contracting era, it is possible to recover the consistency of
the spectral index and of the running index, even in a single
scalar field model \cite{deHaro:2015wda}. However, the
tensor-to-scalar ratio still remains a problem. From a different
perspective, it has recently  been shown \cite{Nojiri:2019lqw}
that an $F(R)$ gravity model with Lagrange multiplier is able to
resolve most of the problems arising in the context of matter or
quasi-matter bounce scenarios, albeit it fails to restore the
energy conditions. In such Lagrange multiplier $F(R)$ gravity
model, the tensor power spectrum becomes smaller in comparison
with that of scalar perturbation and thus the tensor to scalar
ratio lies within Planck constraints, unlike in pure $F(R)$
gravity, where the observable parameters are not simultaneously
compatible with the observational constraints.

The recently observed striking event of neutron star merging GW170817 \cite{GBM:2017lvd}, reflects the
fact that the propagation speed of the gravitational and electromagnetic waves are the same, i.e. equal to one in natural units. This observation
has definitely narrowed down the viability of the modified gravitational theories, since every theory that predicts a gravitational wave
speed different from one is no more a viable description (see Ref.[\cite{Ezquiaga:2017ekz}] for a list of theories that are ruled out by \cite{GBM:2017lvd}).

Motivated by the above arguments, we will here study an extended
matter bounce scenario in the context of a ghost free
$f(R,\mathcal{G})$ model, as developed in \cite{Nojiri:2018ouv},
and the conditions for such  model to have gravitational waves
speed ($c_T^2$) equal to one and thus become consistent with
GW170817. As it was demonstrated in \cite{Nojiri:2018ouv}, the
constraint $c_T^2 = 1$ restricts the form of the Gauss-Bonnet
coupling function. After finding the suitable form which also
realizes the considered bounce model, we address the bouncing
phenomenology of the resulting theory.

The paper is organized as follows. In Sect.[\ref{SecII}], we
briefly discuss the essential features of the ghost free
$f(R,\mathcal{G})$ model compatible with the event GW170817.
Sections [\ref{sec_reconstruction}] and [\ref{sec_perturbation}]
are respectively devoted to an extensive study of the primordial
scalar and tensor perturbations and the observable parameters.
Finally, the conclusions and remarks follow in the end of the
paper.

\section{Essential features of a ghost-free $f(R,\mathcal{G})$ gravity compatible with the GW170817 event\label{SecII}}

In this section we shall recall the essential features of the
ghost free $f(R,\mathcal{G})$ gravity theory developed in Ref.~\cite{Nojiri:2018ouv}. We consider
$f(R,\mathcal{G}) = \frac{R}{2\kappa^2} + f(\mathcal{G})$ which, owing to the presence of $f(\mathcal{G})$, contains ghosts with respect to
 perturbations of the spacetime metric. However the ghost modes may be eliminated by introducing a Lagrange multiplier $\lambda$ in
the standard $f(\mathcal{G})$ gravity action
\cite{Nojiri:2018ouv}, leading to a ghost-free action, as follows
\begin{equation}
\label{FRGBg19} S=\int d^4x\sqrt{-g} \left(\frac{1}{2\kappa^2}R +
\lambda \left( \frac{1}{2} \partial_\mu \chi \partial^\mu \chi +
\frac{\mu^4}{2} \right)
 - \frac{1}{2} \partial_\mu \chi \partial^\mu \chi
+ h\left( \chi \right) \mathcal{G} - V\left( \chi \right) +
\mathcal{L}_\mathrm{matter}\right)\, ,
\end{equation}
where $\mu$ is a constant with mass dimension [+1]. Upon variation
with respect to the Lagrange multiplier $\lambda$, one obtains the
following constraint equation
\begin{equation}
\label{FRGBg20} 0=\frac{1}{2} \partial_\mu \chi \partial^\mu \chi
+ \frac{\mu^4}{2} \, .
\end{equation}
Effectively, the kinetic term is a constant, so it can be absorbed in the scalar potential, as
\begin{equation}
\label{FRGBg21} \tilde V \left(\chi\right) \equiv \frac{1}{2}
\partial_\mu \chi \partial^\mu \chi + V \left( \chi \right)
= - \frac{\mu^4}{2} + V \left( \chi \right) \, ,
\end{equation}
and the action of Eq.~(\ref{FRGBg19}) can be rewritten as
\begin{equation}
\label{FRGBg22} S=\int d^4x\sqrt{-g} \left(\frac{1}{2\kappa^2}R +
\lambda \left( \frac{1}{2} \partial_\mu \chi \partial^\mu \chi +
\frac{\mu^4}{2} \right) + h\left( \chi \right) \mathcal{G}
 - \tilde V\left( \chi \right) + \mathcal{L}_\mathrm{matter}\right)
\, .
\end{equation}
The scalar and gravitational equations of motion for the action (\ref{FRGBg22}) take the  form
\begin{align}
\label{FRGBg23} 0 =& - \frac{1}{\sqrt{-g}} \partial_\mu \left(
\lambda g^{\mu\nu}\sqrt{-g}
\partial_\nu \chi \right)
+ h'\left( \chi \right) \mathcal{G} - {\tilde V}'\left( \chi \right) \, , \\
\label{FRGBg24} 0 =& \frac{1}{2\kappa^2}\left(- R_{\mu\nu} +
\frac{1}{2}g_{\mu\nu} R\right) + \frac{1}{2} T_{\mathrm{matter}\,
\mu\nu}
 - \frac{1}{2} \lambda \partial_\mu \chi \partial_\nu \chi
 - \frac{1}{2}g_{\mu\nu} \tilde V \left( \chi \right)
+ D_{\mu\nu}^{\ \ \tau\eta} \nabla_\tau \nabla_\eta h \left( \chi
\right)\, ,
\end{align}
where $D_{\mu\nu}^{\ \ \tau\eta}$ has the following form,
\begin{eqnarray}
 D_{\mu\nu}^{\ \ \tau\eta}&=&\big(\delta_{\mu}^{\ \tau}\delta_{\nu}^{\ \eta} + \delta_{\nu}^{\ \tau}\delta_{\mu}^{\ \eta} -
 2g_{\mu\nu}g^{\tau\eta}\big)R + \big(-4g^{\rho\tau}\delta_{\mu}^{\ \eta}\delta_{\nu}^{\ \sigma}
 - 4g^{\rho\tau}\delta_{\nu}^{\ \eta}\delta_{\mu}^{\ \sigma} + 4g_{\mu\nu}g^{\rho\tau}g^{\sigma\nu}\big)R_{\rho\sigma}\nonumber\\
 &+&4R_{\mu\nu}g^{\tau\eta} - 2R_{\rho\mu\sigma\nu} \big(g^{\rho\tau}g^{\sigma\nu} + g^{\rho\eta}g^{\sigma\tau}\big)
 \nonumber
\end{eqnarray}
with having in mind $g^{\mu\nu}D_{\mu\nu}^{\ \ \tau\eta} =
4\bigg[-\frac{1}{2}g^{\tau\eta}R + R^{\tau\eta}\bigg]$. Upon
multiplication of Eq.~(\ref{FRGBg24}) with $g^{\mu\nu}$, we get
\begin{equation}
\label{FRGBg24A} 0 = \frac{R}{2\kappa^2} + \frac{1}{2}
T_\mathrm{matter} + \frac{\mu^4}{2} \lambda - 2 \tilde V \left(
\chi \right) - 4 \left( - R^{\tau\eta} + \frac{1}{2} g^{\tau\eta}
R \right) \nabla_\tau \nabla_\eta h \left( \chi \right) \, ,
\end{equation}
and solving  Eq.~(\ref{FRGBg24A}) with respect to $\lambda$ yields
\begin{equation}
\label{FRGBg24AB} \lambda = - \frac{2}{\mu^4} \left(
\frac{R}{2\kappa^2} + \frac{1}{2} T_\mathrm{matter}
 - 2 \tilde V \left( \chi \right) - 4 \left( - R^{\tau\eta}
+ \frac{1}{2} g^{\tau\eta} R \right) \nabla_\tau \nabla_\eta h
\left( \chi \right) \right) \, .
\end{equation}
Eventually, we will deal with non-singular bounce cosmology in
this ghost free $f(R,\mathcal{G})$ model and, thus, the spatially
flat Friedmann-Robertson-Walker (FRW) metric ansatz will fulfill
our purpose. Let us now see how the equations of motion become if
the metric background is a flat FRW one, with line element
\begin{equation}
\label{FRWmetric} ds^2 = - dt^2 + a(t)^2 \sum_{i=1,2,3} \left(
dx^i \right)^2 \, .
\end{equation}
Assuming that the functions $\lambda$ and $\chi$ are only cosmic
time dependent, and also that no matter fluids are present, that
is, that $T_{\mathrm{matter}\, \mu\nu} =0$, then Eq.~(\ref{FRGBg20}) admits the following simple solution
\begin{equation}
\label{frgdS4} \chi = \mu^2 t \, .
\end{equation}
Hence, the $(t,t)$ and $(i,j)$ components of Eq.~(\ref{FRGBg24}) can be written as
\begin{align}
\label{FRGFRW1} 0 = & - \frac{3H^2}{2\kappa^2}
 - \frac{\mu^4 \lambda}{2} + \frac{1}{2} \tilde V \left( \mu^2 t \right)
 - 12 \mu^2 H^3 h' \left( \mu^2 t \right) \, , \\
\label{FRGFRW2} 0 = & \frac{1}{2\kappa^2} \left( 2 \dot H + 3 H^2
\right)
 - \frac{1}{2} \tilde V \left( \mu^2 t \right)
+ 4 \mu^4 H^2 h'' \left( \mu^2 t \right) + 8 \mu^2 \left( \dot H +
H^2 \right) H h' \left( \mu^2 t \right) \, ,
\end{align}
and, in addition, from Eq.~(\ref{FRGBg23}) we get
\begin{equation}
\label{FRGFRW3} 0 = \mu^2 \dot\lambda + 3 \mu^2 H \lambda + 24 H^2
\left( \dot H + H^2 \right) h'\left( \mu^2 t \right)
 - {\tilde V}'\left( \mu^2 t \right) \, .
\end{equation}
By solving Eq.~(\ref{FRGFRW1}) with respect to $\lambda$, we obtain
\begin{equation}
\label{FRGFRW4} \lambda = - \frac{3 H^2}{\mu^4 \kappa^2} +
\frac{1}{\mu^4} \tilde V \left( \mu^2 t \right) - \frac{24}{\mu^2}
H^3 h' \left( \mu^2 t \right) \, .
\end{equation}
It is easy to see that, by combining Eqs.~(\ref{FRGFRW4}) and
(\ref{FRGFRW3}), we obtain Eq.~(\ref{FRGFRW2}). Also by solving
Eq.~(\ref{FRGFRW2}) with respect to the scalar potential $\tilde V
\left( \mu^2 t \right)$, we get
\begin{equation}
\label{FRGFRW7} \tilde V \left( \mu^2 t \right) =
\frac{1}{\kappa^2} \left( 2 \dot H + 3 H^2 \right) + 8 \mu^4 H^2
h'' \left( \mu^2 t \right) + 16 \mu^2 \left( \dot H + H^2 \right)
H h' \left( \mu^2 t \right) \, .
\end{equation}
Hence, for an arbitrarily chosen function $h(\chi (t))$, and with
the potential $\tilde V \left( \chi \right)$ being equal to
\begin{equation}
\label{FRGFRW8} \tilde V \left( \chi \right) = \left[
\frac{1}{\kappa^2} \left( 2 \dot H + 3 H^2 \right) + 8 \mu^4 H^2
h'' \left( \mu^2 t \right) + 16 \mu^2 \left( \dot H + H^2 \right)
H h' \left( \mu^2 t \right) \right]_{t=\frac{\chi}{\mu^2}}\, ,
\end{equation}
then we can realize an arbitrary cosmology corresponding to a
given Hubble rate $H(t)$. Finally, the functional form of the
Lagrange multiplier reads
\begin{equation}
\label{FRGFRW4B} \lambda = \frac{2 \dot H}{\mu^4 \kappa^2} + 8 H^2
h'' \left( \mu^2 t \right) + \frac{8}{\mu^2} \left( 2 \dot H - H^2
\right) H h' \left( \mu^2 t \right) \, .
\end{equation}
As mentioned in the introductory section, here we are interested
on bouncing scenario followed from the theory with Lagrangian
(\ref{FRGBg22}). The cosmological perturbation, in particular the
tensor perturbation caused by the quantum vacuum fluctuation,
indicates the gravitational wave. On other hand, recently the
event GW170817 confirms the detection of the gravitational wave
which may has a different source than the quantum vacuum
fluctuation, and ensures that the speed of the gravitational wave
is equal to the speed of light (in natural units, it is equal to
unity). In the current paper, we consider that all the
gravitational waves generated from different sources (i.e from
black hole merging or from quantum vacuum fluctuations) have same
propagation speed, and due to GW170817, the speed is equal to
unity. With this consideration, we apply the result of GW170817 in
the present context of bouncing scenario. Earlier, the
implications of GW170817 on non-singular bounce has been
investigated, but in a different context i.e from a degenerate
higher order scalar-tensor (DHOST) theory \cite{Ye:2019frg},
unlike to the present paper where we discuss the bouncing scenario
from ghost free $f(R,\mathcal{G})$ gravity model taking the
constraint of GW170817 into account.\\ The resulting theory with
Lagrangian (\ref{FRGBg22}) is a form of the scalar
Einstein-Gauss-Bonnet gravity along with a Lagrange multiplier,
for which it is well known that the speed of gravitational waves
is different from one, in particular, the gravitational wave speed
in the present context can be expressed as
\cite{Hwang:2005hb,Noh:2001ia,Hwang:2002fp},
\begin{eqnarray}
 c_T^2 = 1 + \frac{16\big(\ddot{h}-\dot{h}H\big)}{\frac{1}{\kappa^2} + 16\dot{h}H}
 \label{gravitaional wave speed}
\end{eqnarray}
with $H = \frac{\dot{a}}{a}$ being the Hubble parameter.
Eq.(\ref{gravitaional wave speed}) apparently reflects the non-viability of the model with respect to GW170817 (which validates the fact that
the gravitational and electromagnetic waves have the same propagation speed). However the gravitational wave speed in the ghost
free $f(R,\mathcal{G})$ model becomes one if the coupling function satisfies the following constraint equation \cite{Odintsov:2019clh},
\begin{eqnarray}
 \ddot{h} = \dot{h}H.
 \label{constraint on coupling}
\end{eqnarray}
Thereby, in order to make compatible our present model with
GW170817, we consider such Gauss-Bonnet coupling functions which
obey Eq.(\ref{constraint on coupling}). Thus, a form of the Hubble
parameter (later, we will consider the Hubble parameter in such a
way that it gives an extended scenario of matter bounce cosmology)
fixes in turn the coupling function via Eq.(\ref{constraint on
coupling}). At this stage it is worth mentioning that the new
constraint on $h(\chi)$ also fits with the original equations of
motion,  this being clear from the fact that there are two
independent equations, namely the $(t,t)$ component of the
gravitational equation and the equation for $\chi$, and yet two
unknown functions ($\lambda$(t), $V(\chi)$) to determine. In the
later section, we will consider a Hubble parameter which leads to
an an extended matter bounce cosmology. In this context, it is
worth mentioning that the presence of the potential term
$\tilde{V}(\chi)$ in $f(R,\mathcal{G})$ model to realize a
non-singular bounce cosmology which is also consistent with $c_T^2
= 1$. Let us demonstrate this in more detail by examining the
equations of motion without the potential term. For $\tilde{V} =
0$, eqn.(\ref{FRGFRW2}) takes the following form-
\begin{eqnarray}
\frac{1}{2\kappa^2} \left( 2 \dot H + 3 H^2\right)
+ 4 \mu^4 H^2 h'' \left( \mu^2 t \right) + 8 \mu^2 \left( \dot H + H^2 \right) H h' \left( \mu^2 t \right) = 0
\label{FRW_new1}
\end{eqnarray}
Recall from eqn.(\ref{constraint on coupling}), the Gauss-Bonnet coupling satisfies a constraint equation in order to make the propagation speed
of the gravitational wave unity. On applying this constraint in eqn.(\ref{FRW_new1}), we get,
\begin{eqnarray}
 \big(2\dot{H} + 3H^2\big)~\big(\ddot{h} + \frac{1}{8\kappa^2}\big) = 0
 \label{FRW_new2}
\end{eqnarray}
which immediately leads to the solution as - $\big(2\dot{H} + 3H^2\big) = 0$, OR $\big(\ddot{h} + \frac{1}{8\kappa^2}\big) = 0$. The first
possibility leads to the evolution of the scale factor as $a(t) = t^{2/3}$ i.e an eternal matter dominated universe, On other hand, the second
possibility gives $\ddot{h} = -\frac{1}{8\kappa^2}$ which along with the constraint $\ddot{h} = \dot{h}H$ makes the scale factor evolution as
$a(t) \sim t$ i.e a Milne universe. Thus eqn.(\ref{FRW_new2}) is unable to realize a non-singular bounce cosmology. This indicates that
in absence of potential, $f(R,\mathcal{G})$ model can not realize a non-singular bounce cosmology which is also consistent with
$c_T^2 = 1$. However, in the present paper, as mentioned earlier, we will explore a non-singular bounce scenario in
$f(R,\mathcal{G})$ model compatible with GW170817 and thus the consideration of the potential term becomes obvious from its own right.\\
In the next section, we shall extensively discuss the bouncing scenario of this $f(R,\mathcal{G})$ model, which is both ghost free and
compatible with the GW170817 event.

\section{Realization of the extended matter bounce scenario}\label{sec_reconstruction}

\begin{figure}[!h]
\begin{center}
 \centering
 \includegraphics[width=3.5in,height=2.0in]{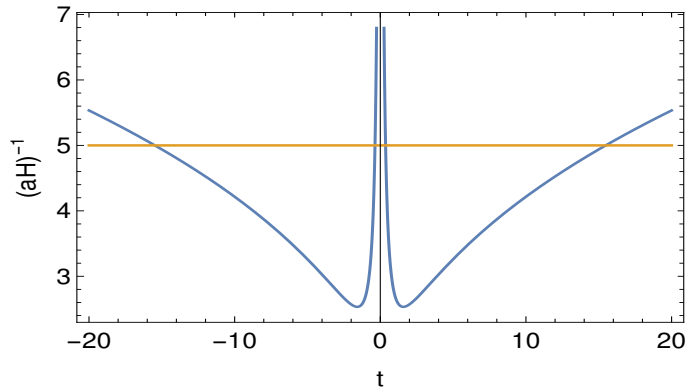}
 \caption{Comoving Hubble radius (Blue curve) and a perturbation mode (yellow curve) with respect to cosmic time for $n = 0.30$}
 \label{plot Hubble radius}
\end{center}
\end{figure}

In this section we will discover which functional
forms for $h(\chi)$ and $\lambda(t)$ can realize a bouncing Universe
cosmological scenario with the following scale factor
\begin{align}
a(t) = \left(a_0t^2 + 1 \right)^n\, ,
\label{scale factor}
\end{align}
where $a_0$ and $n$ are the free parameters of the  model, $a_0$
having mass dimension [+2], while $n$ is dimensionless. The above
scale factor leads to a matter like bouncing universe. Here it may
be mentioned that the scale factor in eqn.(\ref{scale factor}) is
not the unique bouncing model that our universe may realize. There
are other possibilities of bounce, as for example - (1) the
exponential bouncing model where the scale factor evolves as $a(t)
= e^{\alpha t^2}$ and the bounce occurs at $t = 0$, (2) the
singular bounce scenario where the scale facor has the form $a(t)
= e^{\frac{1}{\alpha + 1}(t - t_s)^{\alpha + 1}}$
\cite{Odintsov:2015ynk}. In the case of singular bounce, the
perturbation mode(s) generate near the time of
bouncing unlike to the usual bouncing model where the primordial perturbations generate in the past contracting phase deep inside the Hubble radius.\\
However, as mentioned in the introductory section, the matter-like bounce scenario demonstrated by the scale factor in eqn.(\ref{scale factor}),
faces some questions regarding the observational compatibility with Planck constraints, the instability of linear
order perturbation theory in super horizon scale etc. Motivated by such questions, we investigate the matter-like bounce scenario
in a modified gravity theory like $f(R,G)$ model. Thus the calculations in the following sections of the paper are
based on this choice of the extended matter bounce scale factor. In this sense, one may argue that the results we will obtain depend on the ansatz
of the scale factor shown in eqn.(\ref{scale factor}).\\
The Universe's evolution in a general bouncing cosmology, consists of
two eras, an era of contraction and an era of expansion. The above
scale factor describes a contracting era for the Universe, when
$t\to -\infty$, then the Universe reaches a bouncing point, at
$t=0$, at which the Universe has a minimal size, and then the
Universe starts to expand again, for cosmic times $t>0$. Hence,
the Universe in this scenario never develops a crushing type Big
Bang singularity. It may be mentioned that for $n=1/3$, the scale
factor describes a matter bounce scenario. Eq.~(\ref{scale
factor}) leads to the following Hubble rate and its first
derivative
\begin{align}
H(t) = \frac{2nt}{t^2 + 1/a_0 }\, , \quad
\dot{H}(t) = -2n\frac{t^2 - 1/a_0}{\left(t^2 +1/a_0\right)^2}\, .
\end{align}
With the help of the above expressions, the Ricci scalar is found to be,
\begin{align}
R(t)=12H^2 + 6\dot{H}
=12n \left[\frac{(4n-1)t^2 + 1/a_0}{\left(t^2 + 1/a_0\right)^2}\right] \, .
\label{ricci scalar}
\end{align}
Using Eq.~(\ref{ricci scalar}), one can determine the cosmic time
as a function of the Ricci scalar, that is the function
$t = t(R)$. As a result, the Hubble rate and its first derivative can
be expressed in terms of $R$ (this statement holds true for all
analytic functions of $t$) and, also, the differential operator
$\frac{d}{dt}$ can be written as
$\frac{d}{dt} = \dot{R}(R)\frac{d}{dR}$. However, for the purpose of determining $t = t(R)$ as well as
the observable quantities, we will consider the low curvature limit of the theory. Before proceeding further,
let us comment on the viability of this approximation. We do it in the context of matter bounce cosmology, which is obtained by taking $n=1/3$ in
Eq.~(\ref{scale factor}), the primordial perturbations of the comoving
curvature, which originate from quantum vacuum fluctuations.
At subhorizon scales during the contracting era in the
low-curvature regime,  their wavelength was much smaller
than the comoving Hubble radius, which is defined by $r_h =
\frac{1}{aH}$. In the matter bounce evolution, the Hubble horizon
radius decreases in size, and this causes the perturbation modes
to exit from the horizon, eventually, with this exit occurring when
the contracting Hubble horizon becomes equal to the wavelength of
these primordial modes. However, in the present context, we
consider a larger class of bouncing models of the form $a(t) =
(a_0t^2 + 1)^n$, always within Lagrange
multiplier ghost free Gauss-Bonnet gravity. Thus, it will be important to check what are the
possible values of $n$ which make the low-curvature limit, that is, $R/a_0 \ll 1$ a viable
approximation in calculating the power spectrum for the bouncing
model $a(t) = (a_0t^2 + 1)^n$. This expression of the scale factor immediately leads to the comoving Hubble radius
\begin{align}
r_h = \frac{(1 + a_0t^2)^{1-n}}{2a_0nt} \, .
\label{viability1}
\end{align}
Thereby $r_h$ diverges at $t \simeq 0$, as expected because the
Hubble rate goes to zero at the bouncing point. Furthermore,
the asymptotic behavior of $r_h$ is given by $r_h \sim t^{1-2n}$,
thus $r_h\left(|t|\rightarrow \infty\right)$ diverges for $n < 1/2$,
otherwise $r_h$ goes to zero asymptotically. Hence, for $n < 1/2$,
the comoving Hubble radius decreases initially in the contracting
era and then diverges near the bouncing point; unlike in the case
$n > 1/2$, where the Hubble radius increases from past infinity
and gradually diverges at $t = 0$. As a result, the possible range
of $n$ which leads the perturbation modes to exit the horizon at
large negative time and make the low-curvature limit a viable
approximation in calculating the power spectrum, is given by $0 <
n < 1/2$, see Fig[\ref{plot Hubble radius}], where it is shown that for $n = 0.30$, the perturbation modes can generate inside the Hubble horizon
at a large negative time.
Moreover, we will show in the later sections that this range of $n$ makes the observable quantities compatible
with the Planck constraints and, thus, the ``low curvature limit'' comes as a viable approximation to calculate the power spectra of scalar and tensor
perturbations.\\
In Fig[\ref{plot Hubble radius}], we show the spacetime sketch of
our non-singular bouncing cosmology where the bounce time is taken
at $t = 0$. We also plot the evolution of the physical length
corresponding to a fixed co-moving scale. This scale is the
wavelength of the fluctuation mode $k$, with $k$ standing for the
co-moving wavenumber. The wavelength begins at large negative time
in the contracting phase on sub-Hubble scale, exits the Hubble
radius during this phase at a time (say $-t_h(k)$), and re-enters
the Hubble radius during the low curvature regime in expanding
phase at the time $t_h(k)$ (the exit and entry time are symmetric
about the bouncing point as the scale factor is itself symmetric)
and make the present observation useful. Thus the spacetime
perturbation (or more explicitly the comoving curvature
perturbation) has to be evolved from the contracting phase to the
expanding one, followed by the bouncing phase, in order to get the
power spectrum at late time. In the large scale limit (i.e in the
super-Hubble scale $k \ll aH$) of the contracting phase, the
comoving curvature perturbation ($\Re(k,\tau)$) satisfies the
cosmological perturbation equation
\begin{eqnarray}
 v''(k,\tau) - \frac{z''(\tau)}{z(\tau)}v(k,\tau) = 0
 \label{evolution1}
\end{eqnarray}
where $\tau$ is the conformal time defined as $dt = a(t)d\tau$ and
prime denotes the differentiation with respect to $\tau$
throughout the paper. The above equation is written in terms of
the canonical variable : $v(k,\tau) = z\Re(k,\tau)$, and the
variable $z(\tau)$ depends on the specific model, as for example -
in scalar-tensor theory $z(\tau) = \frac{a\phi'}{H}$ where $\phi$
is the scalar field and for our present considered model (i.e the
Lagrange multiplier ghost free Gauss-Bonnet theory of gravity) the
explicit form of $z(\tau)$ is shown in the later section. However
in terms of a general $z(\tau)$, the solution of
Eq.(\ref{evolution1}) is given by,
\begin{eqnarray}
 v_c(k,\tau) = z(\tau)\bigg[D_{c}(k) + S_c(k)\int^{\tau} \frac{d\tau}{z^2}\bigg]
 \label{evolution2}
\end{eqnarray}
where the suffix 'c' denotes the contracting phase and $D_c(k)$, $S_c(k)$ are independent of time and carry the information about the spectra of the two
modes. The above solution of $v(k,\tau)$ immediately leads to
the curvature perturbation in the super-Hubble scale of the contracting phase as,
\begin{eqnarray}
 \Re_c(k,\tau) = \frac{v(k,\tau)}{z(\tau)} = D_{c}(k) + S_c(k)\int^{\tau} \frac{d\tau}{z^2}
 \label{evolution3}
\end{eqnarray}
As evident from the above expression that the $D$ mode is a
constant mode and generally the $S$ mode appears as an increasing
mode. Similarly in the large scale limit of the expanding phase,
the curvature perturbation has the following solution,
\begin{eqnarray}
 \Re_e(k,\tau) = D_{e}(k) + S_e(k)\int^{\tau} \frac{d\tau}{z^2}
 \label{evolution4}
\end{eqnarray}
The $D_e$ mode of the curvature perturbation is constant in time,
as is the $D_c$ mode in contracting phase. However, the role of
the $S$ mode is very different. In the expanding phase $S_e$ is
the sub-dominant decreasing mode, whereas in the contracting
phase, the behavior gets changed i.e $S_c$ mode is the decreasing
mode. Therefore, the dominant mode of the curvature perturbation
in the period of expansion is $D_e$. This leads to the power
spectrum of the curvature perturbation at late time (which is
useful for the present observation) as
\begin{eqnarray}
 P_e(k) \sim k^3 |D_e(k)|^2
 \label{evolution5}
\end{eqnarray}
At this stage it may be mentioned that a model will be a viable
one if the power spectrum $P_e(k)$ becomes nearly scale invariant
accordance to the observations of Planck 2018. Moreover the
curvature perturbation should be continuous and thus $\Re_c$,
$\Re_e$ have to be matched through the bouncing point as
explicitly performed in \cite{Cai:2008qw,Finelli:2001sr}. During
this matching procedure, the $D_e$ mode may inherit the
contribution from both $D_c$ and $S_c$ modes
\cite{Cai:2008qw,Finelli:2001sr}. However as shown in
\cite{Cai:2008qw}, a nearly scale invariant power spectrum of the
constant mode in the contracting phase ($P_c(k) \sim k^3 |D_c|^2$)
will eventually lead to a scale invariant power spectrum in the
expanding phase. So in the present paper, we consider the bouncing
scenario in Lagrange multiplier Gauss-Bonnet gravity model and
concentrate in
calculating the power spectrum for the constant mode in the contracting phase of the Universe, which will be discussed in detail in the next section.\\
Back to our bouncing model $a(t) = (a_0t^2 + 1)^n$, the Ricci scalar ($R(t)$) during the low-curvature regime (or at large negative time)
can be written as $R(t) \sim \frac{12n(4n-1)}{t^2}$ from
Eq.~(\ref{ricci scalar}). This helps to express the scale factor,
the Hubble rate and its first derivative in terms of  the Ricci scalar $R$, as follows
\begin{align}
a(R) = \frac{\left[12na_0^n(4n-1)\right]^n}{R^n} \, , \quad
H(R) = \pm 2n\sqrt{\frac{R}{12n(4n-1)}} \, , \quad
\dot{H}(R)= -2n\sqrt{\frac{R}{12n(4n-1)}}\, ,
\label{hubble parameter}
\end{align}
the '+' and '-' signs in the expression of $H(R)$ indicate the Hubble parameter in the expanding and contracting phases, respectively. Moreover,
by using $H(t) = 2n/t$ (which is valid in the low curvature regime), we immediately get the coupling function $h(\chi(t))$ from
Eq.(\ref{constraint on coupling}), as
\begin{eqnarray}
 h(t) = h_0 t^{2n+1},
 \label{coupling1}
\end{eqnarray}
where $h_0$ is an integration constant having mass dimension [2n+1].
Further, from the expression $R = 1/t^2$, we obtain the coupling function in terms of the Ricci scalar, as
\begin{eqnarray}
 h(R) = \pm h_0 \frac{\big[12n(4n-1)\big]^{n+\frac{1}{2}}}{R^{n+\frac{1}{2}}}.
 \label{coupling2}
\end{eqnarray}
Again the '+' and '-' signs indicate $h(R)$ in the expanding and
contracting phases, respectively. Plugging back these expressions
of $H(R)$ and $h(R)$ into Eq.(\ref{FRGFRW4B}), we obtain the
explicit form of the Lagrange multiplier $\lambda(R)$ in the low
curvature regime of the contracting phase of the Universe, as
\begin{eqnarray}
 \mu^4\lambda = -\frac{R}{3\kappa^2(4n-1)}\bigg[1 - \frac{8n(2n+1)(n+2)}{\big[12n(4n-1)\big]^{1/2 - n}}\kappa^2h_0R^{1/2 - n}\bigg].
 \label{LM}
\end{eqnarray}
Eqs.(\ref{hubble parameter}), (\ref{coupling2}), and (\ref{LM}), which are all valid in the low curvature approximation, i.e. for
$\frac{R}{a_0} \ll 1$, are the main ingredients to determine the observable quantities (recall that the observable parameters are eventually
evaluated at the time of horizon exit, which in turn occurs at a large negative value of time in the present context) for our
considered $f(R,\mathcal{G})$ model with the scale
factor depicted in Eq.(\ref{scale factor}).

This will be the subject of the next section. Later we will
determine the exact expressions of the Hubble parameter, coupling
function. and Lagrange multiplier, which are valid for all cosmic
time beyond the low curvature approximation.


\section{Cosmological Phenomenology and Viability of the Model}\label{sec_perturbation}

In this section we study the first order metric perturbations of
the theory, following
Refs.~\cite{Hwang:2005hb,Noh:2001ia,Hwang:2002fp}, and where the
scalar and tensor perturbations are calculated for various
variants of the higher curvature models. Scalar, vector and tensor
perturbations are decoupled, as in general relativity, so we can
focus our attention to tensor and scalar perturbations separately.
However before starting the cosmological perturbation
calculations, we first determine the functions $Q$
\cite{Hwang:2005hb,Noh:2001ia,Hwang:2002fp} in the context of the
ghost free $f(R,\mathcal{G})$ gravity model, which will be useful
later. Such functions are defined as
\begin{eqnarray}
 Q_a&=&8\dot{h}H^2 = \frac{32n^2(2n+1)}{\big[12n(4n-1)\big]^{1-n}} h_0R^{1-n}\nonumber\\
 Q_b&=&16\dot{h}H = -\frac{32n(2n+1)}{\big[12n(4n-1)\big]^{1/2 - n}}h_0R^{1/2 - n}\nonumber\\
 Q_c&=&Q_d = 0\nonumber\\
 Q_e&=&32\dot{h}\dot{H} = \frac{64n(2n+1)}{\big[12n(4n-1)\big]^{1-n}} h_0R^{1-n}\nonumber\\
 Q_f&=&-16\big[\ddot{h} - \dot{h}H\big] = 0
 \label{Q_s}
\end{eqnarray}
where we use the forms of $h(t)$ and $H(t)$ that we have derived previously. Moreover the function $Q_f$ becomes zero due to the fact
that the coupling function $h(\chi)$ obeys the constraint Eq.(\ref{constraint on coupling}).

\subsection{Scalar perturbations}

The scalar perturbation of FRW background metric is defined as
follows,
\begin{align}
ds^2 = -(1 + 2\Psi)dt^2 + a(t)^2(1 - 2\Psi)\delta_{ij}dx^{i}dx^{j}\, ,
\label{sp1}
\end{align}
where $\Psi(t,\vec{x})$ denotes the scalar perturbation. In
principle, perturbations should always be expressed in terms of
gauge invariant quantities, in our case the comoving curvature
perturbation defined as $\Re = \Psi - aHv$, where, $v(t,\vec{x})$
is the velocity perturbation. However, we shall work in the
comoving gauge, where the velocity perturbation is taken as zero,
thus with such gauge fixing $\Re = \Psi$. Thereby, we can work
with the perturbed variable $\Psi(t,\vec{x})$. The perturbed
action up to $\Psi^2$ order is \cite{Hwang:2005hb},
\begin{align}
\delta S_{\psi} = \int dt d^3\vec{x} a(t) z(t)^2\left[\dot{\Psi}^2
 - \frac{c_s^2}{a^2}\left(\partial_i\Psi\right)^2\right]\, ,
\label{sp2}
\end{align}
where $z(t)$ and $c_s^2$ (the speed of the scalar perturbation wave) have the following expressions,
\begin{align}
z(t) = \frac{a(t)}{H + \frac{\dot{F} + Q_a}{2F + Q_b}} \sqrt{-\mu^4\lambda + \frac{3Q_a^2}{2F + Q_b} + \frac{Q_aQ_e}{2F + Q_b}}
\label{sp3}
\end{align}
and
\begin{align}
 c_{s}^{2} = \frac{-\mu^4\lambda + \frac{3Q_a^2}{2F + Q_b} + \frac{Q_aQ_e}{2F + Q_b}}
 {-\mu^4\lambda + 3\frac{(\dot{F} + Q_a)^2}{2F + Q_b}}
\label{speed scalar perturbation}
\end{align}
respectively, where $F=1$ in our case. For more details on this we
refer the reader to \cite{Hwang:2005hb}. The definition of the
wave speed is for the general Gauss-Bonnet corrected theory with
$F=\frac{\partial f}{\partial R}$, but in our case $f = R$ and
$F=1$. Also the waves speed is affected from the Gauss-Bonnet
coupling via the functions $Q_f$ and $Q_b$ which in our case have
the form
(\ref{Q}).\\
With the above expressions, we concentrate on determining various
observable quantities and specifically, the spectral index of the
primordial curvature perturbations, the tensor-to-scalar ratio and
the running of the spectral index, which are eventually determined
at the time of horizon exit. For the scale factor we consider in
the present paper, the horizon exit occurs during the
low-curvature regime deeply in the contracting era. Thereby, for
the purpose of finding the observable parameters, the condition
$R/a_0 \ll 1$ stands as a viable approximation.

In the low-curvature limit, we determine various terms present in
the expressions of $z(t)$ and $c_s^2$ (see Eqs.~(\ref{sp3}) and (\ref{speed scalar perturbation})) as,
\begin{align}
\frac{a(t)}{H + \frac{\dot{F} + Q_a}{2F + Q_b}} = \frac{a_0^n}{\tilde{R}^{n + 1/2}}
\frac{1}{2n\bigg[1 - \frac{16n(2n+1)h_0\kappa^2\tilde{R}^{1/2 - n}}{1 - 32n(2n+1)h_0\kappa^2\tilde{R}^{1/2 - n}}\bigg]}
\nonumber
\end{align}
and
\begin{eqnarray}
&-&\mu^4\lambda + \frac{3Q_a^2}{2F + Q_b} + \frac{Q_aQ_e}{2F + Q_b}\nonumber\\
&=&\frac{4n\tilde{R}}{\kappa^2}\bigg[1 - 8n(2n+1)(n+2)h_0\kappa^2\tilde{R}^{1/2 - n} + \frac{768n^3(2n+1)^2h_0^2\kappa^4\tilde{R}^{1 - 2n}}{1 - 32n(2n+1)h_0\kappa^2\tilde{R}^{1/2 - n}}
+ \frac{512n^2(2n+1)^2h_0^2\kappa^4\tilde{R}^{1 - 2n}}{1 - 32n(2n+1)h_0\kappa^2\tilde{R}^{1/2 - n}}\bigg]
\nonumber
\end{eqnarray}
where $\tilde{R} = \frac{R}{12n(4n-1)}$. Consequently $z(t)$ takes the following form,
\begin{align}
z(t) = \frac{a_0^n\big[12n(4n-1)\big]^n}{\kappa R^n}~\frac{\sqrt{P(R)}}{Q(R)}
\label{sp4}
\end{align}
where $P(R)$ and $Q(R)$ are defined as follows,
\begin{align}
P(R) = 4n\bigg[1 - 8n(2n+1)(n+2)h_0\kappa^2\tilde{R}^{1/2 - n} + \frac{768n^3(2n+1)^2h_0^2\kappa^4\tilde{R}^{1 - 2n}}{1 - 32n(2n+1)h_0\kappa^2\tilde{R}^{1/2 - n}}
+ \frac{512n^2(2n+1)^2h_0^2\kappa^4\tilde{R}^{1 - 2n}}{1 - 32n(2n+1)h_0\kappa^2\tilde{R}^{1/2 - n}}\bigg],
\label{P}
\end{align}
and
\begin{align}
Q(R) = 2n\bigg[1 - \frac{16n(2n+1)h_0\kappa^2\tilde{R}^{1/2 - n}}{1 - 32n(2n+1)h_0\kappa^2\tilde{R}^{1/2 - n}}\bigg].
\label{Q}
\end{align}
Moreover $c_s^2$ has the following form
\begin{eqnarray}
 c_s^2 = \frac{\bigg[1 - 8n(2n+1)(n+2)h_0\kappa^2\tilde{R}^{1/2 - n} + \frac{768n^3(2n+1)^2h_0^2\kappa^4\tilde{R}^{1 - 2n}}{1 - 32n(2n+1)h_0\kappa^2\tilde{R}^{1/2 - n}}
+ \frac{512n^2(2n+1)^2h_0^2\kappa^4\tilde{R}^{1 - 2n}}{1 - 32n(2n+1)h_0\kappa^2\tilde{R}^{1/2 - n}}\bigg]}
{\bigg[1 - 8n(2n+1)(n+2)h_0\kappa^2\tilde{R}^{1/2 - n} + \frac{768n^3(2n+1)^2h_0^2\kappa^4\tilde{R}^{1 - 2n}}{1 - 32n(2n+1)h_0\kappa^2\tilde{R}^{1/2 - n}}
\bigg]}
\label{speed}
\end{eqnarray}
Therefore up to first order in $\tilde{R}^{1/2 - n}$, the speed of
the scalar perturbation wave is unity, however here we retain the
terms up to the order of $R^{1-2n}$ and thus $c_s^2$ becomes
different than unity in the low curvature regime of our Universe.
Moreover at the bouncing point, the functions $Q_a = Q_b = 0$ (as
these functions are proportional to the Hubble parameter, see Eq.
(\ref{Q_s})) which lead to the speed of the scalar perturbation
wave $c_s^2 = 1$ at the bouncing phase. The positivity of $c_s^2$
stops the exponential growth of the scalar perturbation near the
bouncing regime, as the squeezing term becomes negligible with
respect to $c_s^2k^2$ in the cosmological perturbation equation,
and makes the first order perturbation calculation reliable.\\
Eq.~(\ref{sp2}) clearly indicates that $\Psi(t,\vec{x})$ is not
canonically normalized and to this end we introduce the well-known
variable as $v = z\Re$ ($= z\Psi$ as we are working in the
comoving gauge). The corresponding fourier mode of the variable
$z$ satisfies,
\begin{align}
\frac{d^2v_k}{d\tau^2} + \left(c_s^2k^2 - \frac{1}{z(\tau)}\frac{d^2z}{d\tau^2}\right)v_k(\tau)
= 0 \, ,
\label{sp5}
\end{align}
where $\tau = \int dt/a(t)$ is the conformal time and $v_k(\tau)$
is the Fourier transformed variable of $v(t,\vec{x})$ for the
$k$th mode. Eq.~(\ref{sp5}) is quite hard to solve analytically in
general, since the function $z$ depends on the background
dynamics and also $c_s^2$ is not constant. However the equation can be solved analytically at super-Hubble scale
(i.e when $c_s^2k^2$ can be neglected with respect to the squeezing term $z''/z$) in the
regime $R/a_0 \ll 1$ as we now show. The conformal time ($\tau$)
is related to the cosmic time ($t$) as $\tau = \int
\frac{dt}{a(t)} = \frac{1}{a_0^n(1-2n)} t^{1-2n}$ for $n \neq
1/2$, however we will show that the observable quantities are
compatible with Planck data \cite{Akrami:2018odb} for $n < 1/2$
and thus we can safely work with the aforementioned expression of
$\tau = \tau(t)$. Using this, we can express the Ricci scalar as a
function of the conformal time,
\begin{align}
R(\tau)= \frac{12n(4n-1)}{t^2}
=\frac{12n(4n-1)}{\left[a_0^n(1-2n)\right]^{2/(1-2n)}}
\frac{1}{\tau^{2/(1-2n)}}\, .
\label{sp6}
\end{align}
Having this in mind, along with Eq.~(\ref{sp4}), we can express
$z$ in terms of $\tau$ as follows,
\begin{align}
z(\tau) = \frac{a_0^n}{\kappa} \left[a_0^n(1-2n)\right]^{\frac{2n}{1-2n}} \frac{\sqrt{P(\tau)}}{Q(\tau)}
\tau^{\frac{2n}{1-2n}}\, .
\label{sp67}
\end{align}
The above expression of $z = z(\tau)$ yields the expression of
$\frac{1}{z}\frac{d^2z}{d\tau^2}$, which is essential for the
cosmological perturbation equation,
\begin{align}
\frac{1}{z}\frac{d^2z}{d\tau^2} = \frac{\xi(\xi-1)}{\tau^2}
\bigg[1 + \frac{4n(2n-1)(1-2n)^2}{(4n-1)\big[12n(4n-1)\big]^{1/2 - n}} \bigg(\frac{R}{a_0}\bigg)^{\frac{1}{2} - n}
+ \frac{16n^2(1-2n)^2(108+36n-n^2)}{(4n-1)\big[12n(4n-1)\big]^{1 - 2n}} \bigg(\frac{R}{a_0}\bigg)^{1 - 2n}\bigg]
\label{sp7}
\end{align}
with $\xi = \frac{(2n)}{(1-2n)}$. In the above expression, we retain the terms up to $(R/a_0)^{1-2n}$ as
the cosmological parameters are eventually determined in the low curvature regime, as discussed earlier. Moreover without any loss of
generality, the integration constant $h_0$ is replaced by $h_0 = \frac{\tilde{h}_0a_0^n}{(2n+1)}$ and $\kappa^2\tilde{h}_0$ is taken as
$\kappa^2\tilde{h}_0 = 1/\sqrt{a_0}$. Recall, the low curvature limit is valid for $n < 1/2$ which clearly
indicate that $\frac{1}{2} - n$ (or $1-2n$) is a positive quantity. Thus the
term within parenthesis in Eq.~(\ref{sp7}) can be safely
considered to be small in the low-curvature regime $R/a_0 \ll 1$.
As a result, $\frac{1}{z}\frac{d^2z}{d\tau^2}$ becomes
proportional to $1/\tau^2$ i.e., $\frac{1}{z}\frac{d^2z}{d\tau^2} =
\sigma/\tau^2$ with,
\begin{eqnarray}
\sigma = \xi(\xi-1)
\bigg[1 + \frac{4n(2n-1)(1-2n)^2}{(4n-1)\big[12n(4n-1)\big]^{1/2 - n}} \bigg(\frac{R}{a_0}\bigg)^{\frac{1}{2} - n}
+ \frac{16n^2(1-2n)^2(108+36n-n^2)}{(4n-1)\big[12n(4n-1)\big]^{1 - 2n}} \bigg(\frac{R}{a_0}\bigg)^{1 - 2n}\bigg]
\label{spnew}
\end{eqnarray}
which is approximately a constant in the era, when the primordial
perturbation modes are generated deeply inside the Hubble radius.
In effect, the cosmological perturbation equation in super-Hubble
scale can be solved as follows,
\begin{eqnarray}
v(k,\tau) \sim \frac{k^{-\nu}}{\sqrt{2}}~\tau^{\frac{1}{2} - \nu} ,
\label{sp8}
\end{eqnarray}
with $\nu = \sqrt{\sigma + \frac{1}{4}}$ and the $k$-dependent
multiplication factor is fixed by assuming the Bunch-Davies vacuum
initially. Having the solution of $v_k(\tau)$ at hand, next we
proceed to evaluate the power spectrum (defined for the
Bunch-Davies vacuum state) corresponding to the $k$-th scalar
perturbation mode, which is defined as follows,
\begin{eqnarray}
P_{\Psi}(k,\tau) = \frac{k^3}{2\pi^2}\left|\Psi_k(\tau)\right|^2
= \frac{k^3}{2\pi^2}\left|\frac{v_k(\tau)}{z(\tau)}\right|^2\, .
\label{sp9}
\end{eqnarray}
In the superhorizon limit, using the mode solution in
Eq.~(\ref{sp8}), we have,
\begin{eqnarray}
P_{\Psi}(k,\tau) = \left[\frac{1}{2\pi}\frac{1}{z|\tau|}\right]^2 \left(\frac{k|\tau|}{2}\right)^{3 - 2\nu}\, .
\label{sp10}
\end{eqnarray}
At this stage it deserves mentioning that in calculating the power
spectrum $P_{\Psi}(k,\tau)$, we use the $linear~order$
perturbation theory, which is reflected through the fact that the
perturbed action is taken up to the quadratic order of the
perturbed variable. However it is important to investigate the
validity of the linear order perturbation in our considered
$f(R,G)$ model. For this purpose, we use the super-horizon
solution of $v(k,\tau)$ from eqn.(\ref{sp8}) and determine the
comoving curvature perturbation in super-Hubble scale as,
\begin{eqnarray}
 \Psi(k,\tau) = \frac{v(k,\tau)}{z(\tau)}
 \label{sp_new1}
\end{eqnarray}
where $z(\tau)$ is given in eqn.(\ref{sp67}). Plugging this expression of $z(\tau)$ in eqn.(\ref{sp_new1}), we get
\begin{eqnarray}
 \Psi(k,\tau) = \bigg(\frac{\kappa}{\sqrt{2}a_0^n\bigg[a_0^n(1-2n)\bigg]^{2n/(1-2n)}}\bigg)~k^{-\nu}\tau^{\frac{1}{2}-\nu-\frac{2n}{1-2n}}
 ~\frac{Q(R)}{\sqrt{P(R)}}
 \label{sp_new2}
\end{eqnarray}
Using eqns.(\ref{P}) and (\ref{Q}), we determine the following expressions of $P(R)$ and $Q(R)$ (in the limit $\kappa^2 R < 1$) as,
\begin{eqnarray}
 P(R) = 4n\bigg[1 - 8n(2n+1)(n+2)h_0\kappa^2\tilde{R}^{1/2 - n} + 384n^3(2n+1)^2h_0^2\kappa^4\tilde{R}^{1 - 2n}
 + 256n^2(2n+1)^2h_0^2\kappa^4\tilde{R}^{1 - 2n}\bigg]\label{sp_new3}
 \end{eqnarray}
 and
 \begin{eqnarray}
 Q(R) = 2n\bigg[1 - 16n(2n+1)h_0\kappa^2\tilde{R}^{1/2 - n} - 512n^2(2n+1)^2h_0^2\kappa^4\tilde{R}^{1 - 2n}\bigg]
 \label{sp_new4}
\end{eqnarray}
respectively. Plugging back the above expressions of $P(R)$ and $Q(R)$ into eqn.(\ref{sp_new2}) yields the perturbed variable $\Psi(k,\tau)$ as,
\begin{eqnarray}
 \Psi(k,\tau) = \frac{\kappa k^{-\nu}\big[a_0^n(1-2n)\big]^{\nu - 1/2}}{\sqrt{2}a_0^n} \bigg[\frac{Z(R)}
 {\tilde{R}^{(\frac{1}{2} - n)(\frac{1}{2} - \nu - \frac{2n}{1-2n})}}\bigg]
 \label{sp_new5}
\end{eqnarray}
where we use $R(\tau) = \frac{12n(4n-1)}{\left[a_0^n(1-2n)\right]^{2/(1-2n)}}\frac{1}{\tau^{2/(1-2n)}}$ from eqn.(\ref{sp6}) and
$Z(R)$ has the following form,
\begin{eqnarray}
 Z(R)&=&\bigg[1 - 4n(2n+1)(2-n)h_0\kappa^2\tilde{R}^{1/2 - n} - 768n^2(2n+1)^2h_0^2\kappa^4\tilde{R}^{1 - 2n}\nonumber\\
 &-&64n^2(2n+1)^2(n+2)h_0^2\kappa^4\tilde{R}^{1 - 2n} - 384n^3(2n+1)^2h_0^2\kappa^4\tilde{R}^{1 - 2n}\bigg]
\end{eqnarray}
The linear order
perturbation theory is valid as long as $k^{3/2} \Psi(k,\tau)$ is less than unity i.e
\begin{eqnarray}
 k^{3/2} \Psi(k,\tau) < 1
 \label{sp_new6}
\end{eqnarray}
In order to find an explicit condition from the above inequality, we consider the exact matter bounce scenario (i.e $n = 1/3$, later
we will show that the exact MBS is consistent with the Planck constraints in our considered model) and put $\nu = 3/2$. With these values of $n$
and $\nu$, eqn.(\ref{sp_new6}) becomes,
\begin{eqnarray}
 \frac{\kappa\sqrt{a_0}}{3\sqrt{2}}~\bigg[1 - \frac{20}{9}\bigg(\frac{\tilde{R}}{a_0}\bigg)^{1/6} -
 \frac{3136}{27}\bigg(\frac{\tilde{R}}{a_0}\bigg)^{1/3}\bigg]\bigg(\frac{\tilde{R}}{a_0}\bigg)^{1/2} < 1
 \label{sp_new7}
\end{eqnarray}
where, as previous, $h_0$ is replaced by $h_0 = \frac{\tilde{h}_0a_0^n}{(2n+1)}$ and $\kappa^2\tilde{h}_0$ is taken as
$\kappa^2\tilde{h}_0 = 1/\sqrt{a_0}$. However the inequality in eqn.(\ref{sp_new7}) is valid in the regime $R > 0$ i.e for entire cosmic time.
This indicates that the linear order perturbation theory is valid in the entire range of cosmic time in the ghost free $F(R,G)$ bouncing model.
However in a scalar-tensor matter bounce model, $P(R)$ and $Q(R)$ take the form as $P = 4/3$ and $Q = 2/3$ respectively, actually
the dependence of Ricci scalar in $P(R)$ and $Q(R)$ (see eqns.(\ref{sp_new3}) and (\ref{sp_new4}))
arise due to the presence of $f(G)$ gravity in the present model. Therefore, in scalar-tensor
matter bounce model, the inequality $k^{3/2} \Psi(k,\tau) < 1$ leads to the following condition (in terms of conformal time) :
\begin{eqnarray}
 |\tau|&>&\frac{3}{18^{1/6}}~\bigg(\frac{\kappa}{a_0}\bigg)^{1/3}\nonumber\\
 &=&\tau_f~~~~ (say)
 \label{sp_new8}
\end{eqnarray}
Eqn.(\ref{sp_new8}) clearly indicates that in a scalar-tensor matter bounce model, the perturbation is less than unity
and makes the linear order perturbation valid in the regime $|\tau| > \tau_f$. Thereby the perturbation theory in a scalar-tensor matter bounce scenario
is ill-defined to address the evolution of the perturbed variable in the entire range of cosmic time. This is
in contrary with our considered $f(R,G)$ theory where the linear order perturbation theory is valid for the entire range of cosmic time, as
explained earlier. Actually the presence of $f(G)$ term in the gravitational action modifies the solution of the comoving curvature perturbation,
which in turn validates the linear order perturbation theory for entire cosmic time.\\
By using Eq.~(\ref{sp10}), we can determine the observable
quantities like spectral index of the primordial curvature
perturbations and the running of spectral index. Before proceeding
to calculate these observable quantities, we will consider first
the tensor power spectrum, which is necessary for evaluating the
tensor-to-scalar ratio.


\subsection{Tensor Perturbations}

Let us now focus on the tensor perturbations, which for the FRW
metric background are defined as
\begin{align}
ds^2 = -dt^2 + a(t)^2\left(\delta_{ij} + h_{ij}\right)dx^idx^j\, ,
\label{tp1}
\end{align}
where $h_{ij}(t,\vec{x})$ is the perturbation. The tensor
perturbation is itself a gauge invariant quantity, and the tensor
perturbed action up to quadratic order is given by
\begin{align}
\delta S_{h} = \int dt d^3\vec{x} a(t) z_T(t)^2\left[\dot{h_{ij}}\dot{h^{ij}}
 - \frac{1}{a^2}\left(\partial_kh_{ij}\right)^2\right]\, ,
\label{tp2}
\end{align}
where $z_T(t)$ and $c_T^2$ (the speed of the gravitational wave) are
\begin{align}
z_T(t) = a\sqrt{F + \frac{1}{2}Q_b}\, .
\label{tp3}
\end{align}
Recall that the coupling function in the present context satisfies
the relation $\ddot{h} = \dot{h}H$, which in turn makes the speed
of the gravitational wave to be one, i.e. $c_T^2 = 1$, and thus
the model becomes compatible with the GW170817 event. Similar to
the case of  scalar perturbations, the variable $z$ for tensor
ones is defined as $(v_T)_{ij} = z_T~h_{ij}$ which, upon
performing the Fourier transformation, satisfies the equation
\begin{align}
\frac{d^2v_T(k,\tau)}{d\tau^2}
+ \left(k^2 - \frac{1}{z_T(\tau)}\frac{d^2z_T}{d\tau^2}\right)v_T(k,\tau)
= 0\, .
\label{tp4}
\end{align}
By using Eq.~(\ref{tp3}), along with the condition $R/a_0 \ll 1$,
we evaluate $z_T(\tau)$ and
$\frac{1}{z_T(\tau)}\frac{d^2z_T}{d\tau^2}$, which read
\begin{align}
z_T(\tau) = \frac{a_0^n}{\kappa} \left[a_0^n(1-2n)\right]^{\frac{2n}{1-2n}} S(\tau) \tau^{\frac{2n}{1-2n}}
\label{tpnew}
\end{align}
and
\begin{align}
\frac{1}{z_T}\frac{d^2z_T}{d\tau^2} = \frac{\xi(\xi-1)}{\tau^2}
\bigg[1 - \frac{16n(1-2n)^2}{(4n-1)\big[12n(4n-1)\big]^{1/2 - n}} \bigg(\frac{R}{a_0}\bigg)^{\frac{1}{2} - n}
- \frac{256n^2(1-2n)^2}{(4n-1)\big[12n(4n-1)\big]^{1 - 2n}} \bigg(\frac{R}{a_0}\bigg)^{1 - 2n}\bigg],
\label{tp6}
\end{align}
respectively, where $S(R(\tau)) = \sqrt{\frac{1}{2}\bigg[1 - \frac{32n(2n+1)}{\big[12n(4n-1)\big]^{1/2 - n}}h_0\kappa^2R^{\frac{1}{2} - n}}$
and also we use $R=R(\tau)$ from Eq.~(\ref{sp6}). Due to the fact that
$n < 1/2$, the variation of the term in
parenthesis in Eq.~(\ref{tp6}) can be regarded to be small in the
low-curvature regime, and thus
$\frac{1}{z_T}\frac{d^2z_T}{d\tau^2}$ becomes proportional to
$1/\tau^2$ that is $\frac{1}{z_T}\frac{d^2z_T}{d\tau^2} =
\sigma_T/\tau^2$, with
\begin{align}
\sigma_T = \xi(\xi-1)
\bigg[1 - \frac{16n(1-2n)^2}{(4n-1)\big[12n(4n-1)\big]^{1/2 - n}} \bigg(\frac{R}{a_0}\bigg)^{\frac{1}{2} - n}
- \frac{256n^2(1-2n)^2}{(4n-1)\big[12n(4n-1)\big]^{1 - 2n}} \bigg(\frac{R}{a_0}\bigg)^{1 - 2n}\bigg],
\label{tp9}
\end{align}
and recall $\xi = \frac{(2n)}{(1-2n)}$. The above expressions
yield the tensor power spectrum, defined with the initial state of
the Bunch-Davies vacuum, so we have,
\begin{align}
P_{h}(k,\tau) = 2\left[\frac{1}{2\pi}\frac{1}{z_T|\tau|}\right]^2 \left(\frac{k|\tau|}{2}
\right)^{3 - 2\nu_T}\, .
\label{tp10}
\end{align}
The factor $2$ arises from the two polarization modes of the
gravity wave, and $\nu_T = \sqrt{\sigma_T + \frac{1}{4}}$, where
$\sigma_T$ is defined in Eq.~(\ref{tp9}).

\begin{figure}[!h]
\begin{center}
 \centering
 \includegraphics[width=3.5in,height=2.0in]{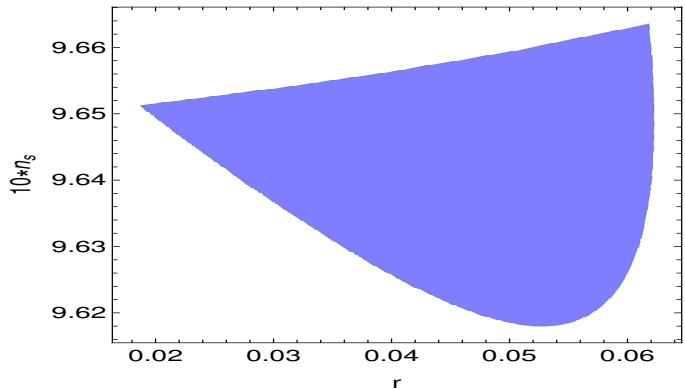}
 \caption{Parametric plot of $10\times n_s$ vs $r$ for
 $0.01 \leq \frac{R_h}{a_0} \leq 0.05$ and $0.30 \lesssim n \lesssim 0.40$.}
 \label{plot1}
\end{center}
\end{figure}

Now we can explicitly confront the model at hand with the latest
Planck observational data \cite{Akrami:2018odb}, so we shall
calculate the spectral index of the primordial curvature
perturbations $n_s$ and the tensor-to-scalar ratio $r$, which are
defined as
\begin{align}
n_s = 1 + \left. \frac{\partial \ln{P_{\Psi}}}{\partial
\ln{k}}\right|_{\tau=\tau_h} \, , \quad
r = \left. \frac{P_{h}(k,\tau)}{P_{\Psi}(k,\tau)}\right|_{\tau=\tau_h}\, .
\label{obs1}
\end{align}
Eqs.~(\ref{sp10}) and (\ref{tp10}) immediately lead to the
explicitly form of $n_s$ and $r$
\begin{align}
n_s = 4 - \sqrt{1 + 4\sigma} \, , \quad
r = 2\left[\frac{z(\tau)}{z_T(\tau)}\right]^2_{\tau = \tau_h}\, ,
\label{obs2}
\end{align}
where $\sigma$, $z(\tau)$ and $z_T(\tau)$ are given in
Eqs.~(\ref{spnew}), (\ref{sp6}), and (\ref{tpnew}), respectively. As it
is evident from the above equations, $n_s$ and $r$ are evaluated
at the time of horizon exit, when $k=aH$, or equivalently at $\tau
= \tau_h$. It should be noticed that $n_s$ and $r$ depend on the
dimensionless parameters $\frac{R_h}{a_0}$ and $n$, with $R_h =
R(\tau_h)$. We can now directly confront the spectral index and
the tensor-to-scalar ratio with the Planck 2018 results
\cite{Akrami:2018odb}, which constrain the observational indices
to be
\begin{equation}
\label{planckconstraints}
n_s = 0.9649 \pm 0.0042\, , \quad r < 0.064\, .
\end{equation}
For the model at hand, $n_s$ and $r$ are within the Planck
constraints for the following ranges of parameter values: $0.01
\leq \frac{R_h}{a_0} \leq 0.05$ and $0.30 \lesssim n \lesssim
0.40$, and this behavior is depicted in Fig.~\ref{plot1}. The
viable range of $R_h/a_0$ is in agreement with the low-curvature
condition $R/a_0 \ll 1$ that we have considered in our
calculations. Moreover, the range of the parameter $n$ clearly
indicates that the matter bounce scenario, for which $n=1/3$, is
well described by the ghost free Einstein-Gauss-Bonnet gravity model with the Lagrange multiplier term. At this stage it is worth mentioning
that Eq.(\ref{spnew}) clearly reveals that in absence of the Gauss-Bonnet coupling term (i.e for $h_0 = 0$, for which the present model resembles
 a scalar-tensor model), the quantity $\sigma$ becomes
$\sigma = 2$ for the pure matter bounce scenario, which in turn makes the power spectrum completely scale invariant (i.e $n_s = 1$).
However, it is well known that
in scalar-tensor theory the matter bounce scenario is not consistent with
the Planck observations. Moreover the matter bounce scenario also does
not fit well even in the standard $F(R)$ gravity, as it was confirmed in a previous paper \cite{Nojiri:2019lqw}.
However, here we show that in the ghost free $f(R,\mathcal{G})$ model which is also compatible with GW170817, the matter bounce
can be considered as a good bouncing model, which allows the
simultaneous compatibility of $n_s$ and $r$ with the astronomical observations.

\begin{figure}[!h]
\begin{center}
 \centering
 \includegraphics[width=3.5in,height=2.0in]{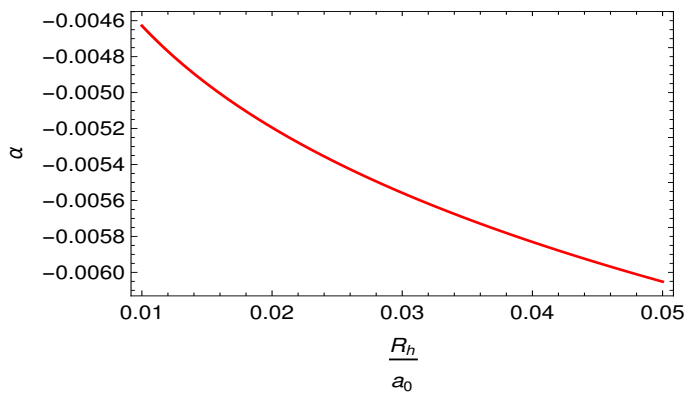}
 \caption{Parametric plot of $\alpha$ vs $\frac{R_h}{a_0}$ for
 $n = 1/3$, i.e. for pure MBS}
 \label{plot2}
\end{center}
\end{figure}

Furthermore, the running of the spectral index is defined
as follows,
\begin{align}
 \alpha = \left. \frac{dn_s}{d\ln{k}} \right|_{\tau=\tau_h}\, ,
 \label{obs3}
\end{align}
and this is constrained by Planck 2018 results as $\alpha =
-0.0085 \pm 0.0073$. Thus, it is also important to calculate the
running of spectral index before concluding the viability of a
model. By using the expression of $\sigma$ (see Eq.~(\ref{spnew}))
and $R = R(\tau)$ (see Eq.~(\ref{sp6})), we get
\begin{align}
\alpha = \frac{8n\xi(\xi-1)(2-n)(1-2n)^3}{\big(12n(4n-1)\big)^{1-2n}\sqrt{1 + 4\xi(\xi-1)}} \bigg(\frac{R_h}{a_0}\bigg)^{\frac{1}{2} - n}
\label{obs4}
\end{align}
To arrive to the above result, we use the horizon crossing
relation of the $k$-th mode $k = aH$ to determine
$\frac{d|\tau|}{d\ln{k}} = -|\tau|$, i.e., the horizon exit time
$|\tau|$ increases as the momentum of the perturbation mode
decreases, as expected. Eq.~(\ref{obs4}) indicates that similarly
to $n_s$ and $r$, the running index ($\alpha$) also depends on the
parameters $R_h/a_0$ and $n$. Taking $n=1/3$, in Fig.~\ref{plot2} we give a
plot of $\alpha$ with respect to $R_h/a_0$ .
As it can be seen in Fig.~\ref{plot2}, the parameter $\alpha$ lies
within the Planck constraint for $0.01 \lesssim R_h/a_0 \lesssim
0.05$, for $n = 1/3$. For the ghost free $f(R,\mathcal{G})$ model,
we have shown that both the pure matter bounce scenario as well as
the quasi-matter bounce scenario are consistent with the Planck
observations, what is not true in the case of scalar-tensor or
pure $F(R)$ models. Therefore, the Lagrange multiplier ghost free
Gauss-Bonnet gravity model has a richer phenomenology as compared
to the scalar-tensor or standard $F(R)$ gravity models, which fail
to describe in a viable way these two {\it bounce cosmology} scenarios.\\
Before proceeding further, it is worth mentioning that in
\cite{Cai:2016thi,Cai:2017tku}, the authors explored how to build
a viable nonsingular cosmological models within the framework of
effective field theory (EFT). They showed that a healthy
nonsingular bounce model can be achieved by introducing an
effective operator of $R^{(3)}\delta g^{00}$, in particular, the
effective action looks like
\begin{eqnarray}
 S_{eff} = \int d^4x \sqrt{-g} &\bigg[&\frac{R}{2\kappa^2} - \Lambda(t) - c(t)g^{00}\nonumber\\
 &+&\frac{M_2^4(t)}{2}\big(\delta g^{00}\big)^2 - \frac{m_3^3(t)}{2}\delta K\delta g^{00} + \frac{\tilde{m}_4^2}{2}R^{(3)}\delta g^{00}\bigg]
 \label{embed1}
\end{eqnarray}
where $\Lambda(t)$, $c(t)$, $M_2(t)$, $m_3(t)$ and $\tilde{m}_4(t)$ are various time dependent coefficients. Having this effective action in hand,
we like to investigate whether the $f(R,\mathcal{G})$ model that we consider in the present paper can be embedded in or have some differences
in respect to this EFT action ($S_{eff}$). The demonstration goes as follows:\\
The first line in the expression (\ref{embed1}) describes the background model and the rest is for perturbations.
Therefore the background Friedmann equations are given by,
\begin{eqnarray}
 \frac{3H^2}{2\kappa^2} = \frac{1}{2}\big[\Lambda(t) + c(t)\big]\nonumber\\
 \frac{1}{2\kappa^2}\big[2\dot{H} + 3H^2\big] = \frac{1}{2}\big[\Lambda(t) - c(t)\big]
 \label{embed2}
\end{eqnarray}
Moreover the effective action (\ref{embed1}) leads to the
quadratic action for scalar and tensor perturbations as \cite{Cai:2016thi},
\begin{eqnarray}
 \delta S_{\Psi} = \int d^4x a^3(t) c_1(t) \bigg[\dot{\Psi}^2 - \frac{1}{c_1}\bigg(\frac{\dot{c}_3}{a} - c_2\bigg)\frac{(\partial \Psi)^2}
 {a^2}\bigg]
 \label{embed3}
\end{eqnarray}
and
\begin{eqnarray}
 \delta S_h = \int d^4x a^3(t) \bigg[\dot{h}_{ij}^2 - \frac{1}{a^2}\big(\partial_k h_{ij}\big)^2\bigg]
 \label{embed4}
\end{eqnarray}
respectively, where $\Psi$ and $h_{ij}$ are scalar and tensor perturbations respectively. Moreover $c_i$ (s) have the following form,
\begin{eqnarray}
 c_1 = \frac{\big[3m_3^6 + 4H^2\epsilon/\kappa^4 + 8M_2^4/\kappa^2\big]}{\kappa^2\big[2H/\kappa^2 - m_3^3\big]^2}~~~~~~~,~~~~~~c_2 = 1/\kappa^2~~~~,~~~~~c_3 = \frac{a\big[1 + 2\kappa^2\tilde{m}_4^2\big]}{\kappa^2\big[H - \kappa^2m_3^3/2\big]}
 \label{embed5}
\end{eqnarray}
Comparing Eqns.(\ref{FRGFRW1}), (\ref{FRGFRW2}) with Eqn.(\ref{embed2}), one can argue that the background equations of our considered
$f(R,\mathcal{G})$ model can be embedded within that of the effective field theory (EFT) action (\ref{embed1}), if the coefficients
$\Lambda(t)$ and $c(t)$ are related with various functions of the $f(R,\mathcal{G})$ model as,
\begin{eqnarray}
 \Lambda(t) = -\frac{\mu^4\lambda}{2} + \tilde{V}(\chi(t)) - 4H^2\ddot{h} - 4H^3\dot{h} + 8H\dot{H}\dot{h}
 \label{embed6}
\end{eqnarray}
and
\begin{eqnarray}
 c(t) = -\frac{\mu^4\lambda}{2} + 4H^2\ddot{h} - 20H^3\dot{h} - 8H\dot{H}\dot{h}
 \label{embed7}
\end{eqnarray}
respectively, recall that $\chi(t) = \mu^2t$. Moreover comparing Eqns.(\ref{sp2}) and (\ref{embed3}), we may conclude that the scalar perturbation of
our considered $f(R,\mathcal{G})$ model is also consistent with that of the effective action (\ref{embed1}), provided that the coefficients $c_i(t)$
are given by,
\begin{eqnarray}
 c_1 = \frac{1}{\bigg[H + \frac{\dot{F} + Q_a}{2F + Q_b}\bigg]^2} \bigg[-\mu^4\lambda + \frac{3Q_a^2}{2F + Q_b} + \frac{Q_aQ_e}{2F + Q_b}\bigg]\nonumber\\
 \frac{1}{c_1}\big(\dot{c}_3 - c_2/a\big) = \frac{-\mu^4\lambda + \frac{3Q_a^2}{1/\kappa^2 + Q_b} + \frac{Q_aQ_e}{1/\kappa^2 + Q_b}}
 {-\mu^4\lambda + \frac{3Q_a^2}{1/\kappa^2 + Q_b}}
 \label{embed8}
\end{eqnarray}
where the functions $Q_i$ are shown in Eqn.(\ref{Q_s}). However eqns.(\ref{tp2}) and (\ref{embed4}) clearly reveal that
the tensor perturbation of the present ghost free $f(R,\mathcal{G})$ model
is not consistent with that of the effective action (\ref{embed1}), because of the time dependence of the factor $z_T$ in eqn.(\ref{tp2}),
unlike in the expression of eqn.(\ref{embed4}). At this stage it deserves mentioning that ``$z_T$'' factor of the tensor perturbation in an
effective field theory may be made time dependent if one adds the terms like $-m_4^2(t)\big(\delta K^2 - \delta K_{\mu\nu}\delta K^{\mu\nu}\big)$ or
$-\tilde{\lambda}(t)\nabla_iR^{(3)}\nabla^{i}R^{(3)}$ to the effective action i.e if $S_{eff}$ takes the following form \cite{Cai:2016thi},
\begin{eqnarray}
 S_{eff} = \int d^4x \sqrt{-g} &\bigg[&\frac{R}{2\kappa^2} - \Lambda(t) - c(t)g^{00}
 + \frac{M_2^4(t)}{2}\big(\delta g^{00}\big)^2 - \frac{m_3^3(t)}{2}\delta K\delta g^{00} + \frac{\tilde{m}_4^2}{2}R^{(3)}\delta g^{00}\nonumber\\
 &-&m_4^2(t)\big(\delta K^2 - \delta K_{\mu\nu}\delta K^{\mu\nu}\big) - \tilde{\lambda}(t)\nabla_iR^{(3)}\nabla^{i}R^{(3)}\bigg]
 \label{embed9}
\end{eqnarray}
The above effective action leads to the quadratic action of tensor perturbation as follows,
\begin{eqnarray}
 \delta S_h = \int d^4x a^3(t)\big(1 + 2\kappa^2m_4^2(t)\big) \bigg[\dot{h}_{ij}^2 - \frac{1}{a^2\big(1 + 2\kappa^2m_4^2(t)\big)}
 \big(\partial_k h_{ij}\big)^2\bigg]
 \label{embed10}
\end{eqnarray}
It is evident from eqn.(\ref{embed10}) that due to the time
dependence of $m_4^2(t)$, the front factor of $\delta S_h$ (apart
from $a^3(t)$) becomes time dependent. Therefore the addition of
the new term like  $-m_4^2(t)\big(\delta K^2 - \delta
K_{\mu\nu}\delta K^{\mu\nu}\big)$ in the effective action makes
the ``$z_T$'' factor of the tensor perturbation time dependent,
which is consistent with our considered $f(R,\mathcal{G})$ model.
However eqn.(\ref{embed10}) also indicates that due to the
presence of the term $-m_4^2(t)\big(\delta K^2 - \delta
K_{\mu\nu}\delta K^{\mu\nu}\big)$, the gravitational wave speed in
the EFT becomes different than unity, unlike to the current
$f(R,\mathcal{G})$ model where the gravitational wave speed is
unity. This in turn clarifies that the present ghost free
$f(R,\mathcal{G})$ model is not consistent to the effective action
(\ref{embed9}). Thus the current Lagrange multiplier Gauss-Bonnet
theory of gravity can neither be embedded within the EFT action
(\ref{embed1}) nor within the action (\ref{embed9}). However with
a non-trivial coefficient (other than unity) of Ricci scalar along
with $m_4 = 0$ in the effective field theory action, in
particular, if the EFT action has the following form
\cite{Cai:2016thi},
\begin{eqnarray}
 S_{eff} = \int d^4x \sqrt{-g} &\bigg[&f(t)\frac{R}{2\kappa^2} - \Lambda(t) - c(t)g^{00}
 + \frac{M_2^4(t)}{2}\big(\delta g^{00}\big)^2 - \frac{m_3^3(t)}{2}\delta K\delta g^{00} + \frac{\tilde{m}_4^2}{2}R^{(3)}\delta g^{00}\bigg]
 \label{embed11}
\end{eqnarray}
then the scalar perturbed action will remain the same as in Eqn.(\ref{embed3}), however the tensor perturbed action comes as,
\begin{eqnarray}
 \delta S_h = \int d^4x a^3(t)f(t) \bigg[\dot{h}_{ij}^2 - \frac{1}{a^2}\big(\partial_k h_{ij}\big)^2\bigg]
 \label{embed12}
\end{eqnarray}
Therefore the effective action in eqn.(\ref{embed11}) makes the front factor of $\delta S_h$ (apart from $a^3(t)$) time dependent and also allow
an unit gravitational wave speed, which are consistent to the present Lagrange multiplier Gauss-Bonnet theory of gravity. Thus the model
considered in the present paper is consistent with the effective action (\ref{embed11}) if the parameters $\Lambda(t)$, $c(t)$ and $f(t)$ satisfy
the following expressions:
\begin{eqnarray}
 \frac{1}{f(t)}\bigg[\frac{1}{2}\big(\Lambda(t) + c(t)\big) - \frac{3\dot{f}H}{2\kappa^2}\bigg] =
 - \frac{\mu^4 \lambda}{2} + \frac{1}{2} \tilde V \left( \mu^2 t \right) - 12 \mu^2 H^3 h' \left( \mu^2 t \right)
 \label{embed13}
\end{eqnarray}
\begin{eqnarray}
 \frac{1}{f(t)}\bigg[\frac{1}{2}\big(\Lambda(t) - c(t)\big) - \frac{2\dot{f}H}{2\kappa^2} - \frac{\ddot{f}}{2\kappa^2}\bigg] =
 \frac{1}{2} \tilde V \left( \mu^2 t \right)
- 4 \mu^4 H^2 h'' \left( \mu^2 t \right) - 8 \mu^2 \left( \dot H +
H^2 \right) H h' \left( \mu^2 t \right)
\label{embed14}
\end{eqnarray}
and
\begin{eqnarray}
 f(t) = \frac{1}{2}\bigg[\frac{1}{\kappa^2} + Q_b\bigg].
 \label{embed15}
\end{eqnarray}
where $Q_b$ is shown in Eq.(\ref{Q_s}). The first two equations
(i.e Eqs.(\ref{embed13}) and (\ref{embed14})) are due to matching
the background equations of motion where as Eq.(\ref{embed15}) are
due to matching the tensor perturbation equation of the EFT
(described by action (\ref{embed11})) with that of the Lagrange
multiplier Gauss-Bonnet theory of gravity.

\subsection{The Energy Conditions Violation Issue for the Bounce}

\begin{figure}[!h]
\begin{center}
 \centering
 \includegraphics[width=3.0in,height=2.0in]{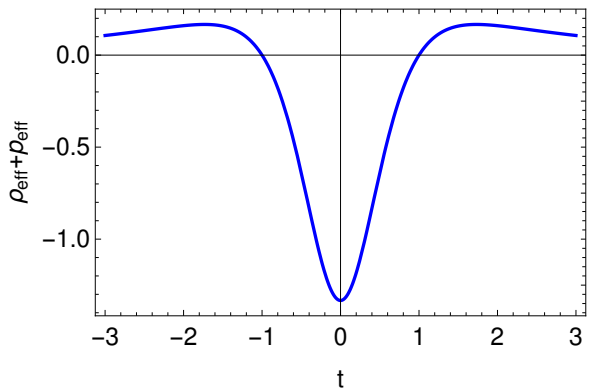}
 \caption{$\rho_\mathrm{eff} + p_\mathrm{eff}$ vs $t$ for the
 purpose of the null energy condition. We take $n = 1/3$, $\tilde{h}_0 = 1$, $a_0 = \frac{1}{\kappa^2} = 1$ (in reduced Planck
units), i.e. we give the plot for the pure matter bounce
scenario.}
 \label{plot_energy1}
\end{center}
\end{figure}

\begin{figure}[!h]
\begin{center}
 \centering
 \includegraphics[width=3.0in,height=2.0in]{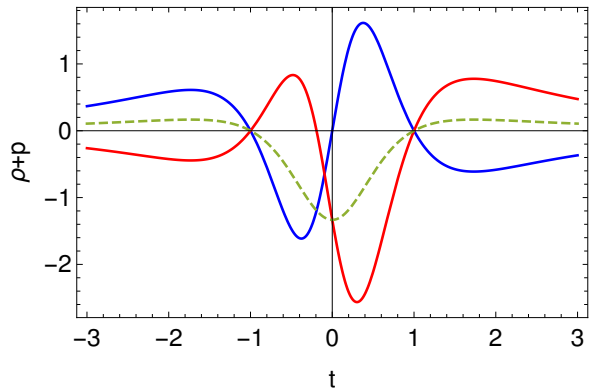}
 \caption{Red Curve: $\big(\rho + p\big)_{\chi}$ vs. $t$. Blue Curve: $\big(\rho + p\big)_{GB}$ vs. $t$. Dashed Curve:
 $\rho_\mathrm{eff} + p_\mathrm{eff}$ vs.$t$. As previously,
 we take $n = 1/3$, $\tilde{h}_0 = 1$, $a_0 = \frac{1}{\kappa^2} = 1$ (in reduced Planck
units), i.e. we give the plot for the pure matter bounce scenario.
For such parametric ranges, $\big(\rho + p\big)_{\chi} =
\frac{4(-1 + t^2)\big(16t + 3(1+t^2)^{2/3}\big)}{9\big(1 +
t^2\big)^{8/3}}$ and $\big(\rho + p\big)_{GB} = \frac{64t(1 -
t^2)}{9\big(1 + t^2\big)^{8/3}}$.}
 \label{plot_energy2}
\end{center}
\end{figure}

At this stage it deserves mentioning that apart from the
computation of spacetime perturbation, the investigation of energy
conditions is also important in a bouncing model from its own
right as most of the bouncing models fail to rescue the null
energy condition. As for example, the energy condition is violated
in the bouncing scenario of Lagrange multiplier $F(R)$ gravity
model \cite{Nojiri:2019lqw}, however the holonomy generalization
of such F(R) model is able to rescue the null energy condition
\cite{Elizalde:2019tee}. Here, we want to investigate the energy
conditions in the present context i.e for the bouncing scenario in
the Lagrange multiplier Gauss-Bonnet theory of gravity. As we
already mentioned, a crucial drawback in most of the bouncing
models is the violation of the null energy condition. Here we will
check the fulfillment of the energy conditions in the context of
the ghost free $f(R,\mathcal{G})$ gravity model. For this purpose,
we first determine the effective energy density $\rho_\mathrm{eff}
+ p_\mathrm{eff}$ from Eqs.~(\ref{FRGFRW1}) and (\ref{FRGFRW2}),
\begin{eqnarray}
 \rho_\mathrm{eff} + p_\mathrm{eff} = -\mu^4\lambda + 8H^2\ddot{h} + 16H\dot{H}\dot{h} - 8H^3\dot{h}
 \label{ec1}
\end{eqnarray}
However, as we mentioned earlier, the energy condition has to be
checked for all cosmic times, including the bouncing point which
occurs at $t=0$, where the low-curvature approximation no longer
holds true. Thus, it will not be justified if we use the forms for
the coupling function and Lagrange multiplier obtained in
Eqs.~(\ref{coupling1}) and (\ref{LM}), to check the energy
condition near the bouncing point. Thereby, the best way to
investigate the energy condition is to determine the forms of
$h(\chi(t))$ and $\lambda(t)$ for the whole range of time
$-\infty < t < \infty$, and then use such forms of $h(t)$ and
$\lambda(t)$ in the expression of $\rho_\mathrm{eff} +
p_\mathrm{eff}$. For this purpose, we reconstruct the GB coupling
function and the Lagrange multiplier term from
Eqs.~(\ref{constraint on coupling}) and (\ref{FRGFRW4B}),
respectively, by using the exact form of $H(t) = \frac{2nt}{t^2 +
1/a_0}$ (which is valid for the entire range of the cosmic time)
\begin{eqnarray}
 \dot{h}(t)&=&\tilde{h}_0 \big(a_0t^2 + 1\big)^n\nonumber\\
 \mu^4\lambda(t)&=&\frac{4a_0n}{\kappa^2}\bigg[\frac{(1-a_0t^2)\big[1 + a_0t\big(t + 16n\tilde{h}_0\kappa^2(t^2 + 1/a_0)^n\big)\big]}
 {\big(1 + a_0t^2\big)^3}\bigg],
 \label{ec2}
\end{eqnarray}
where $\tilde{h}_0$ is an integration constant with mass dimension
[+1], and we should recall that $\tilde{h}_0$ is related to $h_0$
by  $h_0 = \frac{\tilde{h}_0a_0^n}{(2n+1)}$. It should be observed
from the above equation that we stop at the level of the first
derivative of the coupling function and we do not determine the
explicit form of $h(t)$. However, this approximation should be
sufficient, because the right hand side of Eq.(\ref{ec1}) does not
contain any term that is without a derivative of $h(t)$, which is
also expected from the fact that the Gauss-Bonnet term without the
coupling function becomes a surface integral and is eliminated
from the action. Plugging back the expressions of $h(t)$ and
$\lambda(t)$ into Eq.(\ref{ec1}),  in Fig.~\ref{plot_energy1}  we
give the plot of $\rho_\mathrm{eff} + p_\mathrm{eff}$ (with
respect to the cosmic time), for $n = 1/3$, $\tilde{h}_0 = 1$,
$a_0 = \frac{1}{\kappa^2} = 1$ (in reduced Planck units).
As can be seen in Fig.~\ref{plot_energy1}, $\rho_\mathrm{eff} +
p_\mathrm{eff}$ becomes negative near the bouncing point.
Furthermore, Eq.(\ref{ec1}) clearly indicates that both the scalar
field ($\chi$) and the Gauss-Bonnet term contribute to
$\rho_\mathrm{eff} + p_\mathrm{eff}$, their individual
contributions being
\begin{eqnarray}
 \big(\rho + p\big)_{\chi}&=&-\mu^4\lambda\nonumber\\
 &=&\frac{4a_0n}{\kappa^2}\bigg[\frac{(-1 + a_0t^2)\big[1 + a_0t\big(t + 16n\tilde{h}_0\kappa^2(1 + a_0t^2)^n\big)\big]}
 {\big(1 + a_0t^2\big)^3}\bigg]
 \label{ec3}
 \end{eqnarray}
 and
 \begin{eqnarray}
 \big(\rho + p\big)_{GB}&=&8H^2\ddot{h} + 16H\dot{H}\dot{h} - 8H^3\dot{h}\nonumber\\
 &=&\frac{64n^2a_0^2\tilde{h}_0t\big(1 - a_0t^2\big)}{\big(1 + a_0t^2\big)^{3-n}},
 \label{ec4}
\end{eqnarray}
respectively. Using these expressions we give the plot of
$\big(\rho + p\big)_{\chi}$, $\big(\rho + p\big)_{GB}$ and
$\rho_\mathrm{eff} + p_\mathrm{eff}$ (with respect to the cosmic
time, for $n = 1/3$, $\tilde{h}_0 = 1$, $a_0 = \frac{1}{\kappa^2}
= 1$ i.e in reduced Planck units), see Fig. [\ref{plot_energy2}],
in order to better understand the energy flow of the universe.
Figs.[\ref{plot_energy1}] and [\ref{plot_energy2}] show that the
null energy condition for the whole universe, as well as for the
individual contributors, are violated, which further implies that
the weak energy condition is necessarily violated. At this stage,
we want to mention that the holonomy improvement is able to rescue
the energy condition in the context of Lagrange multiplier F(R)
gravity, as proven in \cite{Elizalde:2019tee}. Hopefully,
therefore, the presence of holonomy modifications even in the
$f(R,\mathcal{G})$ gravity model, or the presence of extra spatial
dimensions, where $H^2$ is proportional to linear powers as well
as quadratic powers of the energy density, may play a significant
role to rescue the null energy condition for a non-singular
bounce. This investigation is expected to be carried out soon in a
future work.

\section{Conclusions}\label{sec_conclusion}

In the present paper we have considered an extended matter bounce
scenario in a ghost free $f(R,\mathcal{G})$ gravity, in
particular, for the $f(R,\mathcal{G}) = \frac{R}{2\kappa^2} +
f(\mathcal{G})$ model. The idea for making the model ghost free
consists in introducing a Lagrange multiplier, as discussed in
\cite{Nojiri:2018ouv}. In such gravity theory, we used the results
of \cite{Odintsov:2019clh} which indicated that the model at hand
is compatible with GW170817 that this happens for a class of
Gauss-Bonnet (GB) coupling functions ($h(t)$) which satisfy the
constraint equation $\ddot{h} = \dot{h}H$,  $H(t)$ being the
Hubble parameter. Thus, in order to make our model compatible with
the event GW170817, we considered only  such GB coupling functions
which obey this constraint. At this stage, it is worth mentioning
that this new constraint on the coupling function also fits with
the equations of motion due to the presence of the scalar field
potential.

In such ghost free $f(R,\mathcal{G})$ model compatible with
GW170817, we considered a non-singular bounce scenario with the
scale factor being expressed as $a(t) = (a_0t^2 + 1)^n$, where $n$
is a dimensionless parameter of the model and $t$ the cosmic time.
For this scale factor, it was shown that, for $n < 1/2$, the
spacetime perturbation modes are generated deeply in the
contracting era, at large negative values of time, where the Ricci
curvature is low as compared to the one in the near-bouncing era.
This, in turn, makes the ``low curvature limit''  a viable
approximation in calculating the observable quantities, which are
eventually determined at the time of horizon exit. We have
determined the forms of the coupling function ($h(t)$) and
Lagrange multiplier ($\lambda(t)$) in the low curvature regime, by
using a reconstruction technique for the present model, which
realizes the bouncing with the aforementioned scale factor. Such
forms of $h(t)$ and $\lambda(t)$ led to an explicit expression of
the cosmological perturbation equation, by solving which we have
determined the power spectra of the primordial perturbations and,
correspondingly, have calculated the fundamental cosmological
parameters, namely the spectral index of scalar perturbations, the
tensor to scalar ratio, and the running spectral index,
respectively.

Such observational indexes are found to depend on the
dimensionless parameters of the model, as $R_h/a_0$ (with $R_h$
the Ricci curvature at the time of horizon exit) and $n$. It
turned out that the observable quantities are simultaneously
compatible with the Planck 2018 constraints for the parametric
range $0.01 \lesssim \frac{R_h}{a_0} \lesssim 0.05$ and
$0.30\lesssim n \lesssim 0.40$. It should be noticed that this
range of $n$ is also supported by the range $0 < n < 1/2$, which
ensures the low-curvature approximation to be a perfectly reliable
one for calculating the power spectra. The expression of the
spectral index we obtained clearly demonstrates that, in the
absence of the GB term (the model resembling a scalar tensor one),
the power spectrum becomes completely scale invariant for $n =
1/3$. This was expected, as it is well known that for a scalar
tensor theory, the spectral index becomes one in a pure matter
bounce scenario, which in turn makes the spectrum completely scale
invariant. We recall that the pure matter bounce scenario is not
consistent with observations, even in pure vacuum $F(R)$ gravity.
Here, we showed that for a ghost free $f(R,\mathcal{G})$ model
compatible with the GW170817 event, the matter as well as
quasi-matter bounces may be considered as good bouncing theories,
which confirms the richer bouncing phenomenology of our general
model here. We further determined the effective energy density and
pressure in the present context, in order to investigate the
energy conditions. The energy condition has to be checked for all
cosmic times, including at the bouncing point, where the
low-curvature approximation no longer holds true. Thus, it will
not be justified if we use the forms of coupling function and the
Lagrange multiplier obtained from the low curvature approximation
to check the energy condition near the bouncing point. Keeping
this in mind, we have reconstructed $h(t)$ and $\lambda(t)$ for
the entire range of the cosmic time (by relaxing the low curvature
approximation) and then used such forms of $h(t)$, $\lambda(t)$ to
determine the effective energy density and pressure. As a
consequence, we found that the null energy condition is violated
near the bouncing point, which further implies that the weak
energy condition is necessarily violated. At this stage, we want
to mention that the holonomy improvement has been proven to rescue
the energy conditions in the context of Lagrange multiplier $F(R)$
gravity \cite{Elizalde:2019tee}. Thus, hopefully, the presence of
holonomy modifications, even in the $f(R,\mathcal{G})$ gravity
model, may similarly play a significant role to rescue the null
energy conditions for a non-singular bounce. This investigation
should be interesting and insightful, and we expect to carry it
out in future work, soon.

\section*{Acknowledgments}

EE and SDO acknowledge the support of MINECO (Spain), project FIS2016-76363-P, and  AGAUR (Catalonia, Spain), project 2017 SGR 247.
TP sincerely acknowledges the hospitality by ICE-CSIC/IEEC (Barcelona, Spain), where this work was done during his visit.

\end{document}